# 1. THE COSMOLOGICAL PARAMETERS

Updated November 2013, by O. Lahav (University College London) and A.R. Liddle (University of Edinburgh).

## 1.1. Parametrizing the Universe

Rapid advances in observational cosmology have led to the establishment of a precision cosmological model, with many of the key cosmological parameters determined to one or two significant figure accuracy. Particularly prominent are measurements of cosmic microwave background (CMB) anisotropies, with the highest precision observations being those of the *Planck* Satellite [1,2] which for temperature anisotropies supersede the iconic *WMAP* results [3,4]. However the most accurate model of the Universe requires consideration of a range of different types of observation, with complementary probes providing consistency checks, lifting parameter degeneracies, and enabling the strongest constraints to be placed.

The term 'cosmological parameters' is forever increasing in its scope, and nowadays often includes the parameterization of some functions, as well as simple numbers describing properties of the Universe. The original usage referred to the parameters describing the global dynamics of the Universe, such as its expansion rate and curvature. Also now of great interest is how the matter budget of the Universe is built up from its constituents: baryons, photons, neutrinos, dark matter, and dark energy. We need to describe the nature of perturbations in the Universe, through global statistical descriptors such as the matter and radiation power spectra. There may also be parameters describing the physical state of the Universe, such as the ionization fraction as a function of time during the era since recombination. Typical comparisons of cosmological models with observational data now feature between five and ten parameters.

### 1.1.1. *The global description of the Universe*:

Ordinarily, the Universe is taken to be a perturbed Robertson–Walker space-time with dynamics governed by Einstein's equations. This is described in detail by Olive and Peacock in this volume. Using the density parameters $\Omega_i$ for the various matter species and $\Omega_\Lambda$ for the cosmological constant, the Friedmann equation can be written

$$\sum_i \Omega_i + \Omega_\Lambda - 1 = \frac{k}{R^2 H^2}, \qquad (1.1)$$

where the sum is over all the different species of material in the Universe. This equation applies at any epoch, but later in this article we will use the symbols $\Omega_i$ and $\Omega_\Lambda$ to refer to the present values.

The complete present state of the homogeneous Universe can be described by giving the current values of all the density parameters and the Hubble constant $h$ (the present-day Hubble parameter being written $H_0 = 100 h \,\mathrm{km\,s^{-1}\,Mpc^{-1}}$). A typical collection would be baryons $\Omega_\mathrm{b}$, photons $\Omega_\gamma$, neutrinos $\Omega_\nu$, and cold dark matter $\Omega_\mathrm{c}$ (given charge neutrality, the electron density is guaranteed to be too small to be worth considering separately and is included with the baryons). The spatial curvature can then be determined from the other parameters using Eq. (1.1). The total present matter density $\Omega_\mathrm{m} = \Omega_\mathrm{c} + \Omega_\mathrm{b}$ is sometimes used in place of the cold dark matter density $\Omega_\mathrm{c}$.

These parameters also allow us to track the history of the Universe back in time, at least until an epoch where interactions allow interchanges between the densities of the different species, which is believed to have last happened at neutrino decoupling, shortly before Big Bang Nucleosynthesis (BBN). To probe further back into the Universe's history requires assumptions about particle interactions, and perhaps about the nature of physical laws themselves.

The standard neutrino sector has three flavors. For neutrinos of mass in the range $5 \times 10^{-4}\,\text{eV}$ to $1\,\text{MeV}$, the density parameter in neutrinos is predicted to be

$$\Omega_\nu h^2 = \frac{\sum m_\nu}{93\,\text{eV}}, \tag{1.2}$$

where the sum is over all families with mass in that range (higher masses need a more sophisticated calculation). We use units with $c = 1$ throughout. Results on atmospheric and Solar neutrino oscillations [5] imply non-zero mass-squared differences between the three neutrino flavors. These oscillation experiments cannot tell us the absolute neutrino masses, but within the simple assumption of a mass hierarchy suggest a lower limit of approximately 0.06 eV on the sum of the neutrino masses.

Even a mass this small has a potentially observable effect on the formation of structure, as neutrino free-streaming damps the growth of perturbations. Analyses commonly now either assume a neutrino mass sum fixed at this lower limit, or allow the neutrino mass sum as a variable parameter. To date there is no decisive evidence of any effects from either neutrino masses or an otherwise non-standard neutrino sector, and observations impose quite stringent limits, which we summarize in Section 1.3.4. However, we note that the inclusion of the neutrino mass sum as a free parameter can affect the derived values of other cosmological parameters.

**1.1.2.** *Inflation and perturbations*:

A complete model of the Universe should include a description of deviations from homogeneity, at least in a statistical way. Indeed, some of the most powerful probes of the parameters described above come from the evolution of perturbations, so their study is naturally intertwined in the determination of cosmological parameters.

There are many different notations used to describe the perturbations, both in terms of the quantity used to describe the perturbations and the definition of the statistical measure. We use the dimensionless power spectrum $\Delta^2$ as defined in Olive and Peacock (also denoted $\mathcal{P}$ in some of the literature). If the perturbations obey Gaussian statistics, the power spectrum provides a complete description of their properties.

From a theoretical perspective, a useful quantity to describe the perturbations is the curvature perturbation $\mathcal{R}$, which measures the spatial curvature of a comoving slicing of the space-time. A simple case is the Harrison–Zel'dovich spectrum, which corresponds to a constant $\Delta^2_\mathcal{R}$. More generally, one can approximate the spectrum by a power-law, writing

$$\Delta^2_\mathcal{R}(k) = \Delta^2_\mathcal{R}(k_*) \left[\frac{k}{k_*}\right]^{n_\text{s}-1}, \tag{1.3}$$

where $n_{\rm s}$ is known as the spectral index, always defined so that $n_{\rm s} = 1$ for the Harrison–Zel'dovich spectrum, and $k_*$ is an arbitrarily chosen scale. The initial spectrum, defined at some early epoch of the Universe's history, is usually taken to have a simple form such as this power-law, and we will see that observations require $n_{\rm s}$ close to one. Subsequent evolution will modify the spectrum from its initial form.

The simplest mechanism for generating the observed perturbations is the inflationary cosmology, which posits a period of accelerated expansion in the Universe's early stages [6,7]. It is a useful working hypothesis that this is the sole mechanism for generating perturbations, and it may further be assumed to be the simplest class of inflationary model, where the dynamics are equivalent to that of a single scalar field $\phi$ with canonical kinetic energy slowly rolling on a potential $V(\phi)$. One may seek to verify that this simple picture can match observations and to determine the properties of $V(\phi)$ from the observational data. Alternatively, more complicated models, perhaps motivated by contemporary fundamental physics ideas, may be tested on a model-by-model basis.

Inflation generates perturbations through the amplification of quantum fluctuations, which are stretched to astrophysical scales by the rapid expansion. The simplest models generate two types, density perturbations which come from fluctuations in the scalar field and its corresponding scalar metric perturbation, and gravitational waves which are tensor metric fluctuations. The former experience gravitational instability and lead to structure formation, while the latter can influence the CMB anisotropies. Defining slow-roll parameters, with primes indicating derivatives with respect to the scalar field, as

$$\epsilon = \frac{m_{\rm Pl}^2}{16\pi}\left(\frac{V'}{V}\right)^2 \quad;\quad \eta = \frac{m_{\rm Pl}^2}{8\pi}\frac{V''}{V}, \tag{1.4}$$

which should satisfy $\epsilon, |\eta| \ll 1$, the spectra can be computed using the slow-roll approximation as

$$\Delta_{\mathcal{R}}^2(k) \simeq \left.\frac{8}{3m_{\rm Pl}^4}\frac{V}{\epsilon}\right|_{k=aH} \quad;\quad \Delta_{\rm t}^2(k) \simeq \left.\frac{128}{3m_{\rm Pl}^4}V\right|_{k=aH}. \tag{1.5}$$

In each case, the expressions on the right-hand side are to be evaluated when the scale $k$ is equal to the Hubble radius during inflation. The symbol '$\simeq$' here indicates use of the slow-roll approximation, which is expected to be accurate to a few percent or better.

From these expressions, we can compute the spectral indices [8]

$$n_{\rm s} \simeq 1 - 6\epsilon + 2\eta \quad;\quad n_{\rm t} \simeq -2\epsilon. \tag{1.6}$$

Another useful quantity is the ratio of the two spectra, defined by

$$r \equiv \frac{\Delta_{\rm t}^2(k_*)}{\Delta_{\mathcal{R}}^2(k_*)}. \tag{1.7}$$

We have

$$r \simeq 16\epsilon \simeq -8n_{\rm t}, \tag{1.8}$$

which is known as the consistency equation.

One could consider corrections to the power-law approximation, which we discuss later. However, for now we make the working assumption that the spectra can be approximated by power laws. The consistency equation shows that $r$ and $n_t$ are not independent parameters, and so the simplest inflation models give initial conditions described by three parameters, usually taken as $\Delta_{\mathcal{R}}^2$, $n_s$, and $r$, all to be evaluated at some scale $k_*$, usually the 'statistical center' of the range explored by the data. Alternatively, one could use the parametrization $V$, $\epsilon$, and $\eta$, all evaluated at a point on the putative inflationary potential.

After the perturbations are created in the early Universe, they undergo a complex evolution up until the time they are observed in the present Universe. While the perturbations are small, this can be accurately followed using a linear theory numerical code such as CAMB or CLASS [9]. This works right up to the present for the CMB, but for density perturbations on small scales non-linear evolution is important and can be addressed by a variety of semi-analytical and numerical techniques. However the analysis is made, the outcome of the evolution is in principle determined by the cosmological model, and by the parameters describing the initial perturbations, and hence can be used to determine them.

Of particular interest are CMB anisotropies. Both the total intensity and two independent polarization modes are predicted to have anisotropies. These can be described by the radiation angular power spectra $C_\ell$ as defined in the article of Scott and Smoot in this volume, and again provide a complete description if the density perturbations are Gaussian.

### 1.1.3.  *The standard cosmological model*:

We now have most of the ingredients in place to describe the cosmological model. Beyond those of the previous subsections, we need a measure of the ionization state of the Universe. The Universe is known to be highly ionized at low redshifts (otherwise radiation from distant quasars would be heavily absorbed in the ultra-violet), and the ionized electrons can scatter microwave photons altering the pattern of observed anisotropies. The most convenient parameter to describe this is the optical depth to scattering $\tau$ (*i.e.*, the probability that a given photon scatters once); in the approximation of instantaneous and complete reionization, this could equivalently be described by the redshift of reionization $z_{\rm ion}$.

As described in Sec. 1.4, models based on these parameters are able to give a good fit to the complete set of high-quality data available at present, and indeed some simplification is possible. Observations are consistent with spatial flatness, and indeed the inflation models so far described automatically generate negligible spatial curvature, so we can set $k = 0$; the density parameters then must sum to unity, and so one can be eliminated. The neutrino energy density is often not taken as an independent parameter. Provided the neutrino sector has the standard interactions, the neutrino energy density, while relativistic, can be related to the photon density using thermal physics arguments, and a minimal assumption takes the neutrino mass sum to be that of the lowest mass solution to the neutrino oscillation constraints, namely 0.06 eV. In addition, there is no observational evidence for the existence of tensor perturbations (though the upper limits are fairly weak), and so $r$ could be set to zero. This leaves seven parameters, which is the smallest set that can usefully be compared to the present cosmological data set. This model is referred to by various names, including ΛCDM, the concordance cosmology, and the standard cosmological model.

Of these parameters, only $\Omega_r$ is accurately measured directly. The radiation density is dominated by the energy in the CMB, and the COBE satellite FIRAS experiment determined its temperature to be $T = 2.7255 \pm 0.0006\,\text{K}$ [10], ‡ corresponding to $\Omega_r = 2.47 \times 10^{-5} h^{-2}$. It typically need not be varied in fitting other data. Hence the minimum number of cosmological parameters varied in fits to data is six, though as described below there may additionally be many 'nuisance' parameters necessary to describe astrophysical processes influencing the data.

In addition to this minimal set, there is a range of other parameters which might prove important in future as the data-sets further improve, but for which there is so far no direct evidence, allowing them to be set to a specific value for now. We discuss various speculative options in the next section. For completeness at this point, we mention one other interesting parameter, the helium fraction, which is a non-zero parameter that can affect the CMB anisotropies at a subtle level. Fields, Molaro and Sarkar in this volume discuss current measures of this parameter. It is usually fixed in microwave anisotropy studies, but the data are approaching a level where allowing its variation may become mandatory.

Most attention to date has been on parameter estimation, where a set of parameters is chosen by hand and the aim is to constrain them. Interest has been growing towards the higher-level inference problem of model selection, which compares different choices of parameter sets. Bayesian inference offers an attractive framework for cosmological model selection, setting a tension between model predictiveness and ability to fit the data.

**1.1.4.** *Derived parameters*:

The parameter list of the previous subsection is sufficient to give a complete description of cosmological models which agree with observational data. However, it is not a unique parameterization, and one could instead use parameters derived from that basic set. Parameters which can be obtained from the set given above include the age of the Universe, the present horizon distance, the present neutrino background temperature, the epoch of matter–radiation equality, the epochs of recombination and decoupling, the epoch of transition to an accelerating Universe, the baryon-to-photon ratio, and the baryon to dark matter density ratio. In addition, the physical densities of the matter components, $\Omega_i h^2$, are often more useful than the density parameters. The density perturbation amplitude can be specified in many different ways other than the large-scale primordial amplitude, for instance, in terms of its effect on the CMB, or by specifying a short-scale quantity, a common choice being the present linear-theory mass dispersion on a scale of $8\,h^{-1}\mathrm{Mpc}$, known as $\sigma_8$.

---

‡ Unless stated otherwise, all quoted uncertainties in this article are one-sigma/68% confidence and all upper limits are 95% confidence. Cosmological parameters sometimes have significantly non-Gaussian uncertainties. Throughout we have rounded central values, and especially uncertainties, from original sources in cases where they appear to be given to excessive precision.

Different types of observation are sensitive to different subsets of the full cosmological parameter set, and some are more naturally interpreted in terms of some of the derived parameters of this subsection than on the original base parameter set. In particular, most types of observation feature degeneracies whereby they are unable to separate the effects of simultaneously varying several of the base parameters.

## 1.2. Extensions to the standard model

This section discusses some ways in which the standard model could be extended. At present, there is no positive evidence in favor of any of these possibilities, which are becoming increasingly constrained by the data, though there always remains the possibility of trace effects at a level below present observational capability.

### 1.2.1. *More general perturbations*:

The standard cosmology assumes adiabatic, Gaussian perturbations. Adiabaticity means that all types of material in the Universe share a common perturbation, so that if the space-time is foliated by constant-density hypersurfaces, then all fluids and fields are homogeneous on those slices, with the perturbations completely described by the variation of the spatial curvature of the slices. Gaussianity means that the initial perturbations obey Gaussian statistics, with the amplitudes of waves of different wavenumbers being randomly drawn from a Gaussian distribution of width given by the power spectrum. Note that gravitational instability generates non-Gaussianity; in this context, Gaussianity refers to a property of the initial perturbations, before they evolve.

The simplest inflation models, based on one dynamical field, predict adiabatic perturbations and a level of non-Gaussianity which is too small to be detected by any experiment so far conceived. For present data, the primordial spectra are usually assumed to be power laws.

#### 1.2.1.1. *Non-power-law spectra:*

For typical inflation models, it is an approximation to take the spectra as power laws, albeit usually a good one. As data quality improves, one might expect this approximation to come under pressure, requiring a more accurate description of the initial spectra, particularly for the density perturbations. In general, one can expand $\ln \Delta_{\mathcal{R}}^2$ as

$$\ln \Delta_{\mathcal{R}}^2 (k) = \ln \Delta_{\mathcal{R}}^2 (k_*) + \left(n_{\mathrm{s},*} - 1\right) \ln \frac{k}{k_*} + \frac{1}{2} \left.\frac{dn_{\mathrm{s}}}{d\ln k}\right|_* \ln^2 \frac{k}{k_*} + \cdots, \qquad (1.9)$$

where the coefficients are all evaluated at some scale $k_*$. The term $dn_{\mathrm{s}}/d\ln k|_*$ is often called the running of the spectral index [11]. Once non-power-law spectra are allowed, it is necessary to specify the scale $k_*$ at which the spectral index is defined.

**1.2.1.2.** *Isocurvature perturbations:*

An isocurvature perturbation is one which leaves the total density unperturbed, while perturbing the relative amounts of different materials. If the Universe contains $N$ fluids, there is one growing adiabatic mode and $N-1$ growing isocurvature modes (for reviews see Ref. 12 and Ref. 7). These can be excited, for example, in inflationary models where there are two or more fields which acquire dynamically-important perturbations. If one field decays to form normal matter, while the second survives to become the dark matter, this will generate a cold dark matter isocurvature perturbation.

In general, there are also correlations between the different modes, and so the full set of perturbations is described by a matrix giving the spectra and their correlations. Constraining such a general construct is challenging, though constraints on individual modes are beginning to become meaningful, with no evidence that any other than the adiabatic mode must be non-zero.

**1.2.1.3.** *Seeded perturbations:*

An alternative to laying down perturbations at very early epochs is that they are seeded throughout cosmic history, for instance by topological defects such as cosmic strings. It has long been excluded that these are the sole original of structure, but they could contribute part of the perturbation signal, current limits being just a few percent [13]. In particular, cosmic defects formed in a phase transition ending inflation is a plausible scenario for such a contribution.

**1.2.1.4.** *Non-Gaussianity:*

Multi-field inflation models can also generate primordial non-Gaussianity (reviewed, *e.g.*, in Ref. 7). The extra fields can either be in the same sector of the underlying theory as the inflaton, or completely separate, an interesting example of the latter being the curvaton model [14]. Current upper limits on non-Gaussianity are becoming stringent, but there remains strong motivation to push down those limits and perhaps reveal trace non-Gaussianity in the data. If non-Gaussianity is observed, its nature may favor an inflationary origin, or a different one such as topological defects.

**1.2.2.** *Dark matter properties*:

Dark matter properties are discussed in the article by Drees and Gerbier in this volume. The simplest assumption concerning the dark matter is that it has no significant interactions with other matter, and that its particles have a negligible velocity as far as structure formation is concerned. Such dark matter is described as 'cold,' and candidates include the lightest supersymmetric particle, the axion, and primordial black holes. As far as astrophysicists are concerned, a complete specification of the relevant cold dark matter properties is given by the density parameter $\Omega_c$, though those seeking to directly detect it are as interested in its interaction properties.

Cold dark matter is the standard assumption and gives an excellent fit to observations, except possibly on the shortest scales where there remains some controversy concerning the structure of dwarf galaxies and possible substructure in galaxy halos. It has long been excluded for all the dark matter to have a large velocity dispersion, so-called 'hot' dark matter, as it does not permit galaxies to form; for thermal relics the mass must be above about 1 keV to satisfy this constraint, though relics produced non-thermally, such as the axion, need not obey this limit. However, in future further parameters might need to be introduced to describe dark matter properties relevant to astrophysical observations. Suggestions which have been made include a modest velocity dispersion (warm dark matter) and dark matter self-interactions. There remains the possibility that the dark matter is comprized of two separate components, *e.g.*, a cold one and a hot one, an example being if massive neutrinos have a non-negligible effect.

**1.2.3.    *Relativistic species*:**

The number of relativistic species in the young Universe (omitting photons) is denoted $N_{\rm eff}$. In the standard cosmological model only the three neutrino species contribute, and its baseline value is assumed fixed at 3.046 (the small shift from 3 is because of a slight predicted deviation from a thermal distribution [15]) . However other species could contribute, for example extra neutrino species, possibly of sterile type, or massless Goldstone bosons or other scalars. It is hence interesting to study the effect of allowing this parameter to vary, and indeed although 3.046 is consistent with the data, most analyses currently suggest a somewhat higher value (*e.g.*, Ref. 16).

**1.2.4.    *Dark energy*:**

While the standard cosmological model given above features a cosmological constant, in order to explain observations indicating that the Universe is presently accelerating, further possibilities exist under the general headings of 'dark energy' and 'modified gravity'. These topics are described in detail in the article by Mortonson, Weinberg and White in this volume. This article focuses on the case of the cosmological constant, as this simple case is a good match to existing data. We note that more general treatments of dark energy/modified gravity will lead to weaker constraints on other parameters.

**1.2.5.    *Complex ionization history*:**

The full ionization history of the Universe is given by the ionization fraction as a function of redshift $z$. The simplest scenario takes the ionization to have the small residual value left after recombination up to some redshift $z_{\rm ion}$, at which point the Universe instantaneously reionizes completely. Then there is a one-to-one correspondence between $\tau$ and $z_{\rm ion}$ (that relation, however, also depending on other cosmological parameters). An accurate treatment of this process will track separate histories for hydrogen and helium. While currently rapid ionization appears to be a good approximation, as data improve a more complex ionization history may need to be considered.

**1.2.6.    *Varying 'constants'*:**

Variation of the fundamental constants of Nature over cosmological times is another possible enhancement of the standard cosmology. There is a long history of study of variation of the gravitational constant $G_{\rm N}$, and more recently attention has been drawn to the possibility of small fractional variations in the fine-structure constant. There is presently no observational evidence for the former, which is tightly constrained by a variety of measurements. Evidence for the latter has been claimed from studies of spectral line shifts in quasar spectra at redshift $z \approx 2$ [17], but this is presently controversial and in need of further observational study.

**1.2.7.** *Cosmic topology*:

The usual hypothesis is that the Universe has the simplest topology consistent with its geometry, for example that a flat Universe extends forever. Observations cannot tell us whether that is true, but they can test the possibility of a non-trivial topology on scales up to roughly the present Hubble scale. Extra parameters would be needed to specify both the type and scale of the topology, for example, a cuboidal topology would need specification of the three principal axis lengths. At present, there is no evidence for non-trivial cosmic topology [18].

## 1.3. Probes

The goal of the observational cosmologist is to utilize astronomical information to derive cosmological parameters. The transformation from the observables to the parameters usually involves many assumptions about the nature of the objects, as well as of the dark sector. Below we outline the physical processes involved in each probe, and the main recent results. The first two subsections concern probes of the homogeneous Universe, while the remainder consider constraints from perturbations.

In addition to statistical uncertainties we note three sources of systematic uncertainties that will apply to the cosmological parameters of interest: (i) due to the assumptions on the cosmological model and its priors (*i.e.*, the number of assumed cosmological parameters and their allowed range); (ii) due to the uncertainty in the astrophysics of the objects (*e.g.*, light curve fitting for supernovae or the mass–temperature relation of galaxy clusters); and (iii) due to instrumental and observational limitations (*e.g.*, the effect of 'seeing' on weak gravitational lensing measurements, or beam shape on CMB anisotropy measurements).

These systematics, the last two of which appear as 'nuisance parameters', pose a challenging problem to the statistical analysis. We attempt to fit the whole Universe with 6 to 12 parameters, but we might need to include hundreds of nuisance parameters, some of them highly correlated with the cosmological parameters of interest (for example time-dependent galaxy biasing could mimic growth of mass fluctuations). Fortunately, there is some astrophysical prior knowledge on these effects, and a small number of physically-motivated free parameters would ideally be preferred in the cosmological parameter analysis.

**1.3.1.** *Direct measures of the Hubble constant*:

In 1929, Edwin Hubble discovered the law of expansion of the Universe by measuring distances to nearby galaxies. The slope of the relation between the distance and recession velocity is defined to be the Hubble constant $H_0$. Astronomers argued for decades on the systematic uncertainties in various methods and derived values over the wide range $40 \, \text{km s}^{-1} \, \text{Mpc}^{-1} \lesssim H_0 \lesssim 100 \, \text{km s}^{-1} \, \text{Mpc}^{-1}$.

One of the most reliable results on the Hubble constant comes from the Hubble Space Telescope Key Project [19]. This study used the empirical period–luminosity relations for Cepheid variable stars to obtain distances to 31 galaxies, and calibrated a number of secondary distance indicators—Type Ia Supernovae (SNe Ia), the Tully–Fisher relation, surface-brightness fluctuations, and Type II Supernovae—measured over distances of 400 to 600 Mpc. They estimated $H_0 = 72 \pm 3$ (statistical) $\pm 7$ (systematic) $\mathrm{km\,s^{-1}\,Mpc^{-1}}$.

A recent study [20] of over 600 Cepheids in the host galaxies of eight recent SNe Ia, observed with an improved camera on board the Hubble Space Telescope, was used to calibrate the magnitude–redshift relation for 240 SNe Ia. This yielded an even more precise figure, $H_0 = 73.8 \pm 2.4 \mathrm{\,km\,s^{-1}\,Mpc^{-1}}$ (including both statistical and systematic errors). The major sources of uncertainty in this result are due to the heavy element abundance of the Cepheids and the distance to the fiducial nearby galaxy, the Large Magellanic Cloud, relative to which all Cepheid distances are measured.

The indirect determination of $H_0$ by the *Planck* Collaboration [2] found a lower value, $H_0 = 67.3 \pm 1.2 \mathrm{\,km\,s^{-1}\,Mpc^{-1}}$. As discussed in that paper, there is strong degeneracy of $H_0$ with other parameters, e.g. $\Omega_\mathrm{m}$ and the neutrino mass. The tension between the $H_0$ from *Planck* and the traditional cosmic distance-ladder methods is under investigation.

**1.3.2.  *Supernovae as cosmological probes*:**

Empirically, the peak luminosity of SNe Ia can be used as an efficient distance indicator (*e.g.*, Ref. 21), thus allowing cosmology to be constrained via the distance–redshift relation. The favorite theoretical explanation for SNe Ia is the thermonuclear disruption of carbon–oxygen white dwarfs. Although not perfect 'standard candles', it has been demonstrated that by correcting for a relation between the light curve shape, color, and the luminosity at maximum brightness, the dispersion of the measured luminosities can be greatly reduced. There are several possible systematic effects which may affect the accuracy of the use of SNe Ia as distance indicators, *e.g.*, evolution with redshift and interstellar extinction in the host galaxy and in the Milky Way.

Two major studies, the Supernova Cosmology Project and the High-$z$ Supernova Search Team, found evidence for an accelerating Universe [22], interpreted as due to a cosmological constant or a dark energy component. When combined with the CMB data (which indicates flatness, *i.e.*, $\Omega_\mathrm{m} + \Omega_\Lambda = 1$), the best-fit values were $\Omega_\mathrm{m} \approx 0.3$ and $\Omega_\Lambda \approx 0.7$. Most results in the literature are consistent with the $w = -1$ cosmological constant case. Taking $w = -1$, the SNLS3 team found, by combining their SNIa data with baryon acoustic oscillation (BAO) and *WMAP*7 data, $\Omega_\mathrm{m} = 0.279^{+0.019}_{-0.015}$ and $\Omega_\Lambda = 0.724^{+0.017}_{-0.016}$, including both statistical and systematic errors [23]. This includes a correction for the recently-discovered relationship between host galaxy mass and supernova absolute brightness. This agrees with earlier results [24,25], but note the somewhat higher value for $\Omega_\mathrm{m}$ from *Planck* (see Table 1.1). Future experiments will aim to set constraints on the cosmic equation of state $w(z)$.

1.3.3.  *Cosmic microwave background*:

The physics of the CMB is described in detail by Scott and Smoot in this volume. Before recombination, the baryons and photons are tightly coupled, and the perturbations oscillate in the potential wells generated primarily by the dark matter perturbations. After decoupling, the baryons are free to collapse into those potential wells. The CMB carries a record of conditions at the time of last scattering, often called primary anisotropies. In addition, it is affected by various processes as it propagates towards us, including the effect of a time-varying gravitational potential (the integrated Sachs–Wolfe effect), gravitational lensing, and scattering from ionized gas at low redshift.

The primary anisotropies, the integrated Sachs–Wolfe effect, and scattering from a homogeneous distribution of ionized gas, can all be calculated using linear perturbation theory. Available codes include CAMB and CLASS [9], the former widely used embedded within the analysis package CosmoMC [26]. Gravitational lensing is also calculated in these codes. Secondary effects such as inhomogeneities in the reionization process, and scattering from gravitationally-collapsed gas (the Sunyaev–Zel'dovich (SZ) effect), require more complicated, and more uncertain, calculations.

The upshot is that the detailed pattern of anisotropies depends on all of the cosmological parameters. In a typical cosmology, the anisotropy power spectrum [usually plotted as $\ell(\ell+1)C_\ell$] features a flat plateau at large angular scales (small $\ell$), followed by a series of oscillatory features at higher angular scales, the first and most prominent being at around one degree ($\ell \simeq 200$). These features, known as acoustic peaks, represent the oscillations of the photon–baryon fluid around the time of decoupling. Some features can be closely related to specific parameters—for instance, the location of the first peak probes the spatial geometry, while the relative heights of the peaks probes the baryon density—but many other parameters combine to determine the overall shape.

The 2013 data release from the *Planck* satellite [1] has provided the most powerful results to date on the spectrum of CMB temperature anisotropies, with a precision determination of the temperature power spectrum to beyond $\ell = 2000$, shown in Fig. 1.1. The Atacama Cosmology Telescope (ACT) and South Pole Telescope (SPT) experiments extend these results to higher angular resolution, though without full-sky coverage. The most comprehensive measurements of CMB polarization come from the *WMAP* satellite final (9-year) data release [3], giving the spectrum of *E*-polarization anisotropies and the correlation spectrum between temperature and polarization (those spectra having first been detected by DASI [27]) . These are consistent with models based on the parameters we have described, and provide accurate determinations of many of those parameters [2].

The data provide an exquisite measurement of the location of the first acoustic peak, determining the angular-diameter distance of the last-scattering surface. In combination with other data this strongly constrains the spatial geometry, in a manner consistent with spatial flatness and excluding significantly-curved Universes. CMB data also gives a precision measurement of the age of the Universe. It gives a baryon density consistent with, and at higher precision than, that coming from BBN. It affirms the need for both dark matter and dark energy. It shows no evidence for dynamics of the dark energy, being consistent with a pure cosmological constant ($w = -1$). The density perturbations are consistent with a power-law primordial spectrum, and there is no indication yet of tensor perturbations. The current best-fit for the reionization optical depth from CMB data, $\tau = 0.091$, is in line with models of how early structure formation induces reionization.

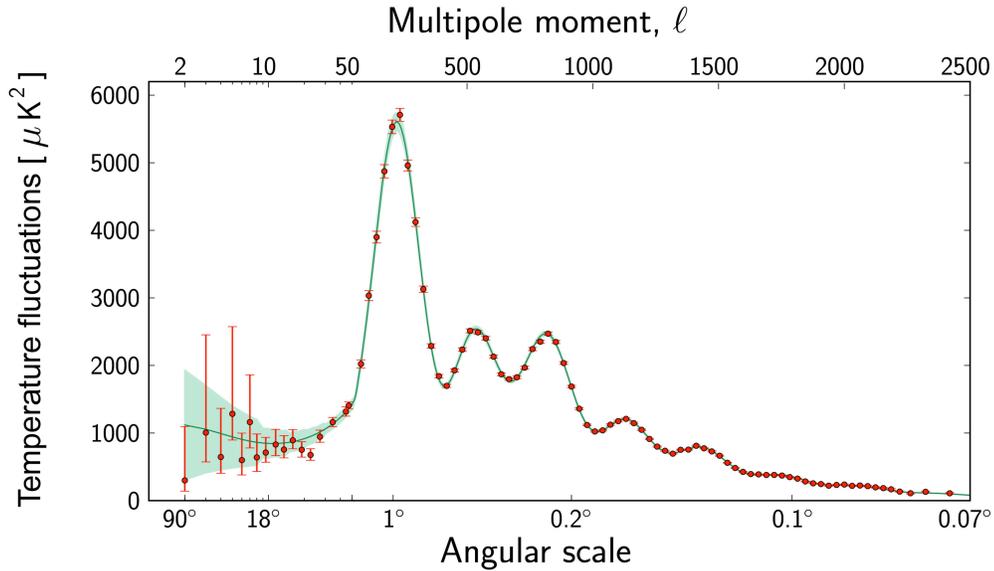

**Figure 1.1:** The angular power spectrum of the CMB temperature anisotropies from *Planck*, from Ref. 1. Note the *x*-axis switches from logarithmic to linear at $\ell = 50$. The solid line shows the prediction from the best-fitting $\Lambda$CDM model and the band indicates the cosmic variance uncertainty.   [Figure courtesy ESA/*Planck* Collaboration.]

*Planck* has also made the first all-sky map of the CMB lensing field, which probes the entire matter distribution in the Universe; this detection corresponds to about $25\sigma$ and adds some additional constraining power to the CMB-only data-sets. ACT previously announced the first detection of gravitational lensing of the CMB from the four-point correlation of temperature variations [28]. These measurements agree with the expected effect in the standard cosmology.

**1.3.4.**  *Galaxy clustering*:

The power spectrum of density perturbations depends on the nature of the dark matter. Within the $\Lambda$CDM model, the power spectrum shape depends primarily on the primordial power spectrum and on the combination $\Omega_{\mathrm{m}}h$ which determines the horizon scale at matter–radiation equality, with a subdominant dependence on the baryon density. The matter distribution is most easily probed by observing the galaxy distribution, but this must be done with care as the galaxies do not perfectly trace the dark matter distribution. Rather, they are a 'biased' tracer of the dark matter. The need to allow for such bias is emphasized by the observation that different types of galaxies show bias with respect to each other. In particular scale-dependent and stochastic biasing may introduce a systematic effect on the determination of cosmological parameters from redshift surveys. Prior knowledge from simulations of galaxy formation or from gravitational lensing data could help to quantify biasing. Furthermore, the observed 3D galaxy distribution is in redshift space, *i.e.*, the observed redshift is the sum of the Hubble expansion and the line-of-sight peculiar velocity, leading to linear and non-linear dynamical effects which also depend on the cosmological parameters. On the largest length scales, the galaxies are expected to trace the location of the dark matter, except for a constant multiplier $b$ to the power spectrum, known as the linear bias parameter. On scales smaller than 20 $h^{-1}$ Mpc or so, the clustering pattern is 'squashed' in the radial direction due to coherent infall, which depends approximately on the parameter $\beta \equiv \Omega_m^{0.6}/b$ (on these shorter scales, more complicated forms of biasing are not excluded by the data). On scales of a few $h^{-1}$ Mpc, there is an effect of elongation along the line of sight (colloquially known as the 'finger of God' effect) which depends on the galaxy velocity dispersion.

**1.3.4.1.** *Baryonic acoustic oscillations:*

The power spectra of the 2-degree Field (2dF) Galaxy Redshift Survey and the Sloan Digital Sky Survey (SDSS) are well fit by a ΛCDM model and both surveys showed evidence for BAOs [29,30]. The Baryon Oscillation Spectroscopic Survey (BOSS) of Luminous Red Galaxies (LRGs) in the SDSS found consistency with the dark energy equation of state $w = -1$ to within $\pm 0.06$ [31]. Similar results for $w$ were obtained by the WiggleZ survey [32]. The BAO data from recent galaxy redshift surveys together with SN Ia data are shown in a Hubble diagram in Figure 1.2.

**1.3.4.2.** *Redshift distortion:*

There is renewed interest in the 'redshift distortion' effect. As the measured redshift of a galaxy is the sum of its redshift due to the Hubble expansion and its peculiar velocity, this distortion depends on cosmological parameters [34] via the perturbation growth rate in linear theory $f(z) = d \ln \delta / d \ln a \approx \Omega^{\gamma}(z)$, where $\gamma \simeq 0.55$ for the ΛCDM model and is different for modified gravity models. Recent observational results show that by measuring $f(z)$ it is feasible to constrain $\gamma$ and rule out certain modified gravity models [35,36]. We note the degeneracy of the redshift-distortion pattern and the geometric distortion (the so-called Alcock–Paczynski effect), *e.g.* as illustrated by the WiggleZ survey [37].

**1.3.4.3.** *Integrated Sachs–Wolfe effect:*

The integrated Sachs–Wolfe (ISW) effect, described in the article by Scott and Smoot, is the change in CMB photon energy when propagating through the changing gravitational potential wells of developing cosmic structures. In linear theory, the ISW signal is expected in universes where there is dark energy, curvature, or modified gravity. Correlating the large-angle CMB anisotropies with very large scale structures, first proposed in Ref. 38, has provided results which vary from no detection of this effect to $4\sigma$ detection [39,40].

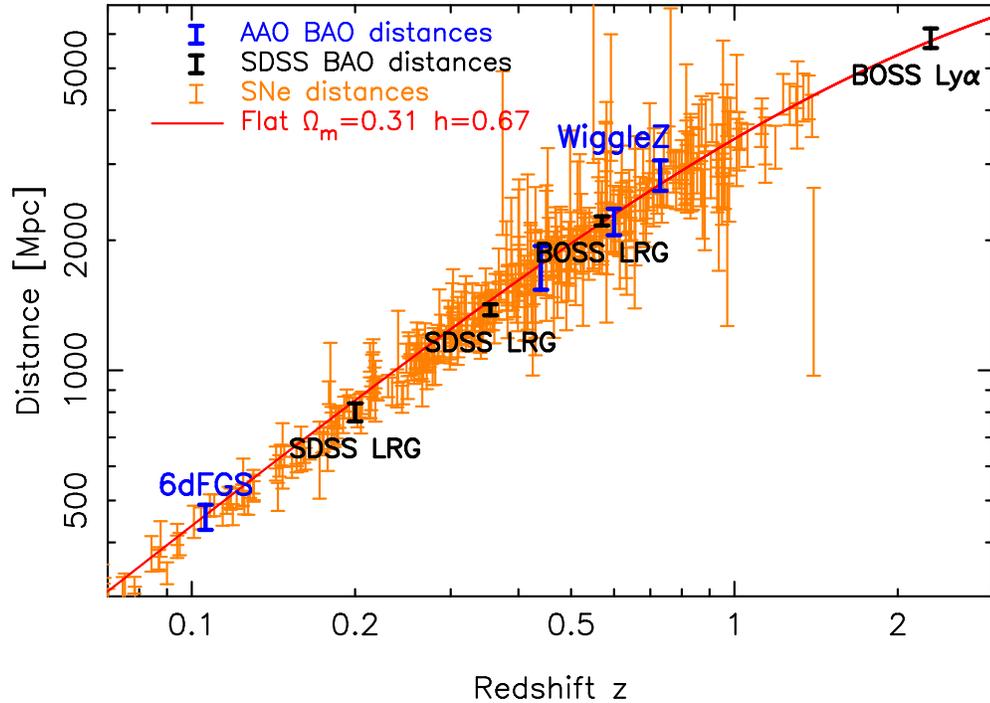

**Figure 1.2:** The cosmic distance scale with redshift. This modern version of the 'Hubble Diagram' combines data from SN Ia as standard candles and BAO as standard rulers in the LRG SSDS, BOSS, 6dFGRS, and WiggleZ galaxy surveys and from the BOSS Lyman-alpha at high redshift. [Figure courtesy of C. Blake, based on Ref. 33.]

**1.3.4.4.**  *Limits on neutrino mass from galaxy surveys and other probes:*

Large-scale structure data constraints on $\Omega_\nu$ due to the neutrino free-streaming effect [41]. Presently there is no clear detection, and upper limits on neutrino mass are commonly estimated by comparing the observed galaxy power spectrum with a four-component model of baryons, cold dark matter, a cosmological constant, and massive neutrinos. Such analyses also assume that the primordial power spectrum is adiabatic, scale-invariant, and Gaussian. Potential systematic effects include biasing of the galaxy distribution and non-linearities of the power spectrum. An upper limit can also be derived from CMB anisotropies alone, while additional cosmological data-sets can improve the results.

Results using a photometric redshift sample of LRGs combined with *WMAP*, BAO, Hubble constant and SNe Ia data gave a 95% confidence upper limit on the total neutrino mass of 0.28eV [42]. Recent spectroscopic redshift surveys, with more accurate redshifts but fewer galaxies, yielded similar upper limits for assumed flat ΛCDM model and additional data-sets: 0.34eV from BOSS [43] and 0.29eV from WiggleZ [44]. *Planck* + *WMAP* polarization + highL CMB [2] give an upper limit of 0.66eV, and with additional BAO data 0.23eV. The effective number of relativistic degrees of freedom is $N_{\rm eff} = 3.30 \pm 0.27$ in good agreement with the standard value $N_{\rm eff} = 3.046$. While the latest cosmological data do not yet constrain the sum of neutrino masses to below 0.2eV, as the lower limit on neutrino mass from terrestrial experiments is 0.06eV, it looks promising that future cosmological surveys will detect the neutrino mass.

### 1.3.5. *Clusters of galaxies*:

A cluster of galaxies is a large collection of galaxies held together by their mutual gravitational attraction. The largest ones are around $10^{15}$ Solar masses, and are the largest gravitationally-collapsed structures in the Universe. Even at the present epoch they are relatively rare, with only a few percent of galaxies being in clusters. They provide various ways to study the cosmological parameters.

The first objects of a given kind form at the rare high peaks of the density distribution, and if the primordial density perturbations are Gaussian distributed, their number density is exponentially sensitive to the size of the perturbations, and hence can strongly constrain it. Clusters are an ideal application in the present Universe. They are usually used to constrain the amplitude $\sigma_8$, as a sphere of radius $8\,h^{-1}$ Mpc contains about the right amount of material to form a cluster. One of the most useful observations at present are of X-ray emission from hot gas lying within the cluster, whose temperature is typically a few keV, and which can be used to estimate the mass of the cluster. A theoretical prediction for the mass function of clusters can come either from semi-analytic arguments or from numerical simulations. The same approach can be adopted at high redshift (which for clusters means redshifts of order one) to attempt to measure $\sigma_8$ at an earlier epoch. The evolution of $\sigma_8$ is primarily driven by the value of the matter density $\Omega_{\rm m}$, with a sub-dominant dependence on the dark energy properties.

The *Planck* observations were used to produce a sample of 189 clusters selected by the SZ effect. The cluster mass function was constructed using a relation between the SZ signal $Y$ and cluster mass $M$. For an assumed flat $\Lambda$CDM model, the *Planck* Collaboration found $\sigma_8 = 0.77 \pm 0.02$ and $\Omega_{\rm m} = 0.29 \pm 0.02$ [45]. Somewhat larger values of both parameters are preferred by the *Planck*'s measurements of the primary CMB anisotropies. The discrepancy might be resolved, for example, by using a different $Y$–$M$ calibration. For comparison with other results in the literature see their Fig. 10.

### 1.3.6. *Clustering in the inter-galactic medium*:

It is commonly assumed, based on hydrodynamic simulations, that the neutral hydrogen in the inter-galactic medium (IGM) can be related to the underlying mass distribution. It is then possible to estimate the matter power spectrum on scales of a few megaparsecs from the absorption observed in quasar spectra, the so-called Lyman-$\alpha$ forest. The usual procedure is to measure the power spectrum of the transmitted flux, and then to infer the mass power spectrum. Photo-ionization heating by the ultraviolet background radiation and adiabatic cooling by the expansion of the Universe combine to give a simple power-law relation between the gas temperature and the baryon density. It also follows that there is a power-law relation between the optical depth $\tau$ and $\rho_{\rm b}$. Therefore, the observed flux $F = \exp(-\tau)$ is strongly correlated with $\rho_{\rm b}$, which itself traces the mass density. The matter and flux power spectra can be related by

$$P_{\rm m}\,(k) = b^2\,(k)\,P_{\rm F}\,(k)\,, \tag{1.10}$$

where $b(k)$ is a bias function which is calibrated from simulations. The BOSS survey has been used to detect and measure the BAO feature in the Lyman-$\alpha$ forest fluctuation at redshift $z = 2.4$, with a result impressively consistent with the standard $\Lambda$CDM model [46]. The Lyman-$\alpha$ flux power spectrum has also been used to constrain the nature of dark matter, for example constraining the amount of warm dark matter [47].

**1.3.7.    *Gravitational lensing*:**

Images of background galaxies are distorted by the gravitational effect of mass variations along the line of sight. Deep gravitational potential wells such as galaxy clusters generate 'strong lensing', leading to arcs, arclets and multiple images, while more moderate perturbations give rise to 'weak lensing'. Weak lensing is now widely used to measure the mass power spectrum in selected regions of the sky (see Ref. 48 for reviews). As the signal is weak, the image of deformed galaxy shapes (the 'shear map') must be analyzed statistically to measure the power spectrum, higher moments, and cosmological parameters.

The shear measurements are mainly sensitive to a combination of $\Omega_\mathrm{m}$ and the amplitude $\sigma_8$. For example, the weak-lensing signal detected by the CFHTLens Survey (over 154 sq. deg. in 5 optical bands) yields, for a flat $\Lambda$CDM model, $\sigma_8(\Omega_\mathrm{m}/0.27)^{0.6} = 0.79 \pm 0.03$ [49]. Earlier results for comparison are summarized in Ref. 48. There are various systematic effects in the interpretation of weak lensing, *e.g.*, due to atmospheric distortions during observations, the redshift distribution of the background galaxies, the intrinsic correlation of galaxy shapes, and non-linear modeling uncertainties.

**1.3.8.    *Peculiar velocities*:**

Deviations from the Hubble flow directly probe the mass perturbations in the Universe, and hence provide a powerful probe of the dark matter [50]. Peculiar velocities are deduced from the difference between the redshift and the distance of a galaxy. The observational difficulty is in accurately measuring distances to galaxies. Even the best distance indicators (*e.g.*, the Tully–Fisher relation) give an uncertainty of 15% per galaxy, hence limiting the application of the method at large distances. Peculiar velocities are mainly sensitive to $\Omega_\mathrm{m}$, not to $\Omega_\Lambda$ or dark energy. While at present cosmological parameters derived from peculiar velocities are strongly affected by random and systematic errors, a new generation of surveys may improve their accuracy. Three promising approaches are the 6dF near-infrared survey of 15,000 peculiar velocities, peculiar velocities of SNe Ia, and the kinematic Sunyaev–Zel'dovich effect.

## 1.4. Bringing observations together

Although it contains two ingredients—dark matter and dark energy—which have not yet been verified by laboratory experiments, the $\Lambda$CDM model is almost universally accepted by cosmologists as the best description of the present data. The approximate values of some of the key parameters are $\Omega_\mathrm{b} \approx 0.05$, $\Omega_\mathrm{c} \approx 0.25$, $\Omega_\Lambda \approx 0.70$, and a Hubble constant $h \approx 0.70$. The spatial geometry is very close to flat (and usually assumed to be precisely flat), and the initial perturbations Gaussian, adiabatic, and nearly scale-invariant.

The most powerful data source is the CMB, which on its own supports all these main tenets. Values for some parameters, as given in Ade *et al.* [2] and Hinshaw *et al.* [4], are reproduced in Table 1.1. These particular results presume a flat Universe. The constraints are somewhat strengthened by adding additional data-sets such as BAO, as shown in the Table, though most of the constraining power resides in the CMB data. We see that the *Planck* and *WMAP* constraints are similar, though with some shifts within the uncertainties. For these six-parameter fits the parameter uncertainties are also comparable; the additional precision of *Planck* data versus *WMAP* is only really apparent when considering significantly larger parameter sets.

If the assumption of spatial flatness is lifted, it turns out that the CMB on its own only weakly constrains the spatial curvature, due to a parameter degeneracy in the angular-diameter distance. However inclusion of other data readily removes this. For example, inclusion of BAO data, plus the assumption that the dark energy is a cosmological constant, yields a constraint on $\Omega_{\rm tot} \equiv \sum \Omega_i + \Omega_\Lambda$ of $\Omega_{\rm tot} = 1.0005 \pm 0.0033$ [2]. Results of this type are normally taken as justifying the restriction to flat cosmologies.

One parameter which is very robust is the age of the Universe, as there is a useful coincidence that for a flat Universe the position of the first peak is strongly correlated with the age. The CMB data give $13.81 \pm 0.05$ Gyr (assuming flatness). This is in good agreement with the ages of the oldest globular clusters and radioactive dating.

The baryon density $\Omega_{\rm b}$ is now measured with high accuracy from CMB data alone, and is consistent with the determination from BBN; Fields *et al.* in this volume quote the range $0.021 \leq \Omega_{\rm b} h^2 \leq 0.025$ (95% confidence).

While $\Omega_\Lambda$ is measured to be non-zero with very high confidence, there is no evidence of evolution of the dark energy density. Mortonson *et al.* in this volume quote the constraint $w = -1.13^{+0.13}_{-0.11}$ on a constant equation of state from a compilation of CMB and BAO data, with the cosmological constant case $w = -1$ giving an excellent fit to the data. Allowing more complicated forms of dark energy weakens the limits.

The data provide strong support for the main predictions of the simplest inflation models: spatial flatness and adiabatic, Gaussian, nearly scale-invariant density perturbations. But it is disappointing that there is no sign of primordial gravitational waves, with the CMB data compilation providing an upper limit $r < 0.11$ at 95% confidence [2] (weakening to 0.26 if running is allowed). The spectral index is clearly required to be less than one by this data, though the strength of that conclusion can weaken if additional parameters are included in the model fits.

Tests have been made for various types of non-Gaussianity, a particular example being a parameter $f_{\rm NL}$ which measures a quadratic contribution to the perturbations. Various non-gaussianity shapes are possible (see Ref. 51 for details), and current constraints on the popular 'local', 'equilateral', and 'orthogonal' types are $f_{\rm NL}^{\rm local} = 3 \pm 6$, $f_{\rm NL}^{\rm equil} = -42 \pm 75$, and $f_{\rm NL}^{\rm ortho} = -25 \pm 39$ (these look weak, but prominent non-Gaussianity requires the product $f_{\rm NL} \Delta_{\mathcal{R}}$ to be large, and $\Delta_{\mathcal{R}}$ is of order $10^{-5}$). Clearly none of these give any indication of primordial non-gaussianity.

**Table 1.1:** Parameter constraints reproduced from Ref. 2 (Table 5) and Ref. 4 (Table 4), with some additional rounding. All columns assume the $\Lambda$CDM cosmology with a power-law initial spectrum, no tensors, spatial flatness, and a cosmological constant as dark energy. *Planck* take the sum of neutrino masses fixed to 0.06eV, while *WMAP* set it to zero. Above the line are the six parameter combinations actually fit to the data in the *Planck* analysis ($\theta_{\rm MC}$ is a measure of the sound horizon at last scattering); those below the line are derived from these. Two different data combinations including *Planck* are shown to highlight the extent to which additional data improve constraints. The first column is a combination of CMB data only — *Planck* temperature plus *WMAP* polarization data plus high-resolution data from ACT and SPT — while the second column adds BAO data from the SDSS, BOSS, 6dF, and WiggleZ surveys. For comparison the last column shows the final nine-year results from the *WMAP* satellite, combined with the same BAO data and high-resolution CMB data (which they call eCMB). Note *WMAP* use $\Omega_\Lambda$ directly as a fit parameter rather than $\theta_{\rm MC}$. The perturbation amplitude $\Delta_\mathcal{R}^2$ is specified at the scale $0.05\,{\rm Mpc}^{-1}$ for *Planck*, but $0.002\,{\rm Mpc}^{-1}$ for *WMAP*, so the spectral index $n_{\rm s}$ needs to be taken into account in comparing them. Uncertainties are shown at 68% confidence.

|  | *Planck*+WP +highL | *Planck*+WP +highL+BAO | *WMAP*9+eCMB +BAO |
|---|---|---|---|
| $\Omega_{\rm b} h^2$ | $0.02207 \pm 0.00027$ | $0.02214 \pm 0.00024$ | $0.02211 \pm 0.00034$ |
| $\Omega_{\rm c} h^2$ | $0.1198 \pm 0.0026$ | $0.1187 \pm 0.0017$ | $0.1162 \pm 0.0020$ |
| $100\,\theta_{\rm MC}$ | $1.0413 \pm 0.0006$ | $1.0415 \pm 0.0006$ | – |
| $n_{\rm s}$ | $0.958 \pm 0.007$ | $0.961 \pm 0.005$ | $0.958 \pm 0.008$ |
| $\tau$ | $0.091^{+0.013}_{-0.014}$ | $0.092 \pm 0.013$ | $0.079^{+0.011}_{-0.012}$ |
| $\ln(10^{10}\Delta_\mathcal{R}^2)$ | $3.090 \pm 0.025$ | $3.091 \pm 0.025$ | $3.212 \pm 0.029$ |
| $h$ | $0.673 \pm 0.012$ | $0.678 \pm 0.008$ | $0.688 \pm 0.008$ |
| $\sigma_8$ | $0.828 \pm 0.012$ | $0.826 \pm 0.012$ | $0.822^{+0.013}_{-0.014}$ |
| $\Omega_{\rm m}$ | $0.315^{+0.016}_{-0.017}$ | $0.308 \pm 0.010$ | $0.293 \pm 0.010$ |
| $\Omega_\Lambda$ | $0.685^{+0.017}_{-0.016}$ | $0.692 \pm 0.010$ | $0.707 \pm 0.010$ |

## 1.5. Outlook for the future

The concordance model is now well established, and there seems little room left for any dramatic revision of this paradigm. A measure of the strength of that statement is how difficult it has proven to formulate convincing alternatives.

Should there indeed be no major revision of the current paradigm, we can expect future developments to take one of two directions. Either the existing parameter set will continue to prove sufficient to explain the data, with the parameters subject to ever-tightening constraints, or it will become necessary to deploy new parameters. The latter outcome would be very much the more interesting, offering a route towards understanding new physical processes relevant to the cosmological evolution. There are many possibilities on offer for striking discoveries, for example:

- The cosmological effects of a neutrino mass may be unambiguously detected, shedding light on fundamental neutrino properties;
- Detection of primordial non-Gaussianities would indicate that non-linear processes influence the perturbation generation mechanism;
- Detection of variation in the dark-energy density (*i.e.*, $w \neq -1$) would provide much-needed experimental input into the nature of the properties of the dark energy.

These provide more than enough motivation for continued efforts to test the cosmological model and improve its accuracy.

Over the coming years, there are a wide range of new observations which will bring further precision to cosmological studies. Indeed, there are far too many for us to be able to mention them all here, and so we will just highlight a few areas.

The CMB observations will improve in several directions. A current frontier is the study of polarization, first detected in 2002 by DASI and for which power spectrum measurements have now been made by several experiments. *Planck* will announce its first polarization results in 2014. Future measurements may be able to separately detect the two modes of polarization and a number of projects are underway with this goal.

An impressive array of dark energy surveys are already operational, under construction, or proposed, including ground-based imaging surveys the Dark Energy Survey and LSST, spectroscopic surveys such as MS-DESI, and space missions *Euclid* and *WFIRST*.

An exciting area for the future is radio surveys of the redshifted 21-cm line of hydrogen. Because of the intrinsic narrowness of this line, by tuning the bandpass the emission from narrow redshift slices of the Universe will be measured to extremely high redshift, probing the details of the reionization process at redshifts up to perhaps 20. LOFAR is the first instrument able to do this and is beginning its operations. In the longer term, the Square Kilometre Array (SKA) will take these studies to a precision level.

The development of the first precision cosmological model is a major achievement. However, it is important not to lose sight of the motivation for developing such a model, which is to understand the underlying physical processes at work governing the Universe's evolution. On that side, progress has been much less dramatic. For instance, there are many proposals for the nature of the dark matter, but no consensus as to which is correct. The nature of the dark energy remains a mystery. Even the baryon density, now measured to an accuracy of a percent, lacks an underlying theory able to predict it within orders of magnitude. Precision cosmology may have arrived, but at present many key questions remain to motivate and challenge the cosmology community.